\documentclass[12pt,a4paper,onecolumn,pra,nofootinbib,showkeys,showpacs,preprint,amsmath,amssymb]{revtex4}

\usepackage[french,german,british]{babel}
\usepackage{amsthm}
\usepackage[dvipdf]{graphicx}
\usepackage[a4paper]{geometry}
\usepackage{bm}
\usepackage{url}

\def\ket#1{|\,#1\,\rangle}
\def\bra#1{\langle\, #1\,|}
\def\braket#1#2{\langle\, #1\,|\,#2\,\rangle}

\allowdisplaybreaks[1]

\begin{document}

\bibliographystyle{prl}

\title{Quantum mechanics as ``space-time statistical mechanics''?}

\author{Anders M\aa nsson}
\email{andman@imit.kth.se}

\affiliation{Institutionen f{\"o}r mikroelektronik och
  informationsteknik,\\
Kungliga Tekniska H{\"o}gskolan\\
Isafjordsgatan 22, SE-164\,40 Kista, Sweden}

\date{\today}

\begin{abstract}
In this paper we discuss and analyse the idea of trying to see
(non-relativistic) quantum mechanics as a ``space-time statistical
mechanics'', by using the classical statistical mechanical method on
objective microscopic space-time configurations. It is argued that this
could perhaps be accomplished by giving up the assumption that the 
objective ``state'' of a system is independent of a future measurement 
performed on the system. This idea is then applied in an example of
quantum state estimation on a qubit system.
\end{abstract}

\keywords{retrocausation, space-time, statistical mechanics, quantum
  state estimation}

\pacs{01.70.+w, 03.65.Ca, 03.65.Ta}
\maketitle

\section{Introduction}
\label{sec:introduction}

The intention of this article is to analyse some aspects of quantum 
mechanics in its standard form and try to see them from a space-time perspective. 
By doing so, perhaps one could get ideas for a new interpretation 
or a new conceptual foundation of quantum mechanics. Things that are 
hard to understand or seem strange in the standard 
form of quantum mechanics, could perhaps be understood in a better and more
clear way if we see quantum mechanics from a space-time
perspective. In this paper we are not trying 
or claiming to give a complete description or explanation of such
things as what the elementary constituents of nature are, which laws 
govern their dynamics, etc. The basic idea presented in this paper is,
in short, that there are objective configurations corresponding to a 
system between preparation and measurement and that these
configurations are determined by both preparation \text{and}
measurement observables. In sections II-IV background and motivation
for this idea is presented. Section V introduces the idea of 
``space-time statistical mechanics'' and in section VI this idea is 
applied in an example of quantum state estimation on a qubit system.     

\section{The quantum state}
\label{sec:qmf}
Let us consider a spin-1/2 system of an electron that has 
been prepared in the quantum state $\ket{\!z+\!}$. 
Formally, the quantum state \textit{before} the
measurement, in this case $\ket{\!z+\!}$, is independent of the 
choice of measurement observable. But, that the system and 
measurement apparatus can be treated
independently in the theory, does not mean they are also independent in
objective reality. (The concept of objective reality is something
idealised and does not necessarily exist, but it is probably fair to say
that to many physicists the final goal of science is to discover and describe the 
objective reality behind what we experience. Doing so, it would be
important to separate theory from the concept of objective reality. 
There are often many different models that are able to account for the same
experimental facts and data. Thus in the creation of a theory 
one has a freedom of choice when it comes to designing the theory.)\\

Consider a classical two level system isolated from its
environment. In classical mechanics one assumes that the two levels are objective 
configurations of the system and that the system always is in one or
the other of those configurations. If the exact configuration of the system
is unknown, one could describe the system as a statistical
distribution $\{p,(1-p)\}$ over the two configurations. A preparation
could then be a procedure that puts the system in such a distribution 
and the preparation would in this sense completely 
determine the statistical ``state'' of the system. But this is not how an 
isolated two level system, prepared (in the
quantum mechanical meaning of the word preparation) in some way, is
described in quantum mechanics. Instead it is mathematically described 
as a \textit{set} of statistical distributions, e.g.
\begin{equation}\label{eq1}
\begin{split}
\ket{\!z+\!} &=
\big\{\{p_{(z+)(\alpha+)},p_{(z+)(\alpha-)}\},\{p_{(z+)(\beta+)},p_{(z+)(\beta-)}\},...\big\}\\
&= \big\{\{|\braket{\!\alpha\!+\!\!\!}{\!z\!+\!}|^2,|\braket{\!\alpha\!\!-\!\!\!}{\!z\!+\!}|^2\},
\{|\braket{\!\beta\!+\!\!\!}{\!z\!+\!}|^2,|\braket{\!\beta\!\!-\!\!\!}{\!z\!+\!}|^2\},...\big\},
\end{split}
\end{equation}
where every distribution corresponds to an observable\footnote{For
  simplicity we do not concern ourselves with POVMs and consider
  only pure states.} that can be
measured, $\{\alpha,\beta,..\}$, and the probabilities for a 
specific ensemble are given by the usual
rules for calculating probabilities in quantum mechanics. There is of course
nothing peculiar or signs of any strange physics by describing the system in this
way, as this only lists every possible statistical distribution corresponding to a 
particular experimental arrangement measuring some observable.
But, strictly seen, in quantum mechanics one only speaks about preparations and outcomes of
measurements performed on a system \cite{peres95}. Quantum mechanics
does not really speak 
about objective configurations of a system\footnote{However, this 
  depends on the interpretation of quantum mechanics one considers.} and one runs into
difficulties when trying to replace the description of the system in 
\eqref{eq1} with only one statistical distribution over some objective configurations. 
There have been attempts to create microscopic models (hidden
variables) behind QM, as for
example the interpretation of QM by Bohm (a combined particle-wave
model) \cite{bohm52I, bohm52II}. These type of models often feels somewhat constructed and too
fantastic to be true. A reasonable approach would be to first see if it
is not possible to explain QM with more established concepts, for example
the particle and space-time concepts without postulating such things as
guiding waves or fields as in Bohm's interpretation.

\section{The wave-particle duality}

The contents of this section are probably well known to the reader, but will be repeated
for clarity. Assume we have a single photon source that can emit single photons on
demand. When we press the button on the source, the single photon detector
clicks (with some delay time of course, due to the photon traveling time). (See figure 1.)\\

\begin{figure}
\begin{center}
\includegraphics[scale=1]{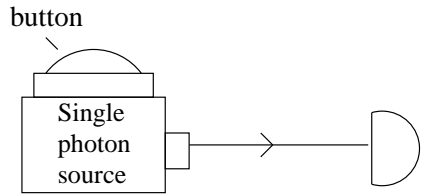}\\
\caption{A single-photon-source and a single-photon-detector.}
\end{center}
\end{figure}

\begin{figure}
\begin{center}
\includegraphics[scale=1]{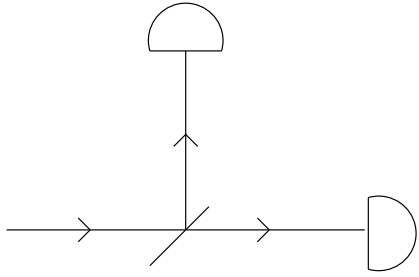}\\
\caption{A 50/50 beamsplitter and two single-photon-detectors.}
\end{center}
\end{figure}

\begin{figure}
\begin{center}
\includegraphics[scale=1]{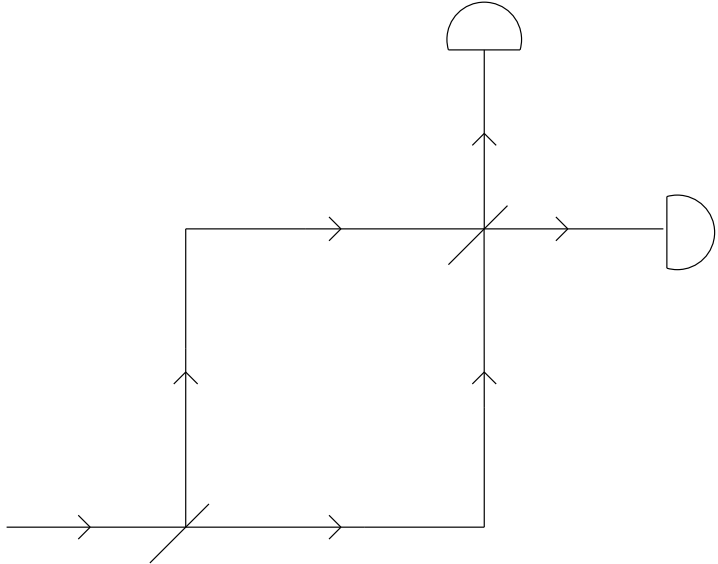}\\
\caption{Two identical 50/50 BS and two single-photon-detectors.}
\end{center}
\end{figure}

If we place a 50/50 beamsplitter (BS) in the photon's path, as in 
figure 2, there will be one and only one click in one of the detectors,
and there is a 50\% probability for either one of the detectors to
click. If one of the detectors clicks, one knows with certainty 
that the other detector will not click. After a click in one of the 
detectors there are no more detectable traces of the photon anywhere.
The fact that only one of the detectors clicks is hard to explain 
if the photon is of purely ``wave nature''. If the photon where of
purely ``wave nature'', one would expect the wave packet to split
into two waves when passing through the BS and there
would be no reason why only one of the detectors would click and the
other one not. This seems to suggest that whatever the photon is, it is something
localized in one and only one of the arms. That is, at the
beamsplitter, the photon goes out in one and only one of the arms. In
this sense the photon seems rather to be of ``particle nature''.\\

But what happens if we let the two photon paths meet again at
another BS, as in figure 3? (We assume all the time that there is only one photon
present, i.e. we only push the single-photon-source button once and do
not push it again until the first photon is detected.) Since the
photon seemed to be localized in only one of the arms and the BS
are identical, if we repeat the experiment several times, we would in
this case expect each of the detectors to click 50\% of the time.
But what actually happens when one performs the
experiment is that one of the detectors always clicks and the
other one never clicks! This is not what we would expect if the photon
is of purely ``particle nature'', i.e. localized in one and only one of the 
arms behind the first BS. But in the setup in figure 3 
we could explain that one of the detectors always and
the other one never clicks, if the photon is of ``wave nature''. But we have
already seen and argued in the setup with only one BS that a ``wave
nature'' of the photon is not able to account for the behaviour with that 
setup. So the photon seems to be of neither purely wave nature nor purely 
particle nature.\\

But maybe the behaviour of the photon depends on the whole
experimental setup in some way? To look into this question, consider 
again the setup in figure 3. But this time, while the photon is on its 
way to the detectors, we have the additional choice to detect the
photon behind the first BS instead (see figure 4). The dotted line
indicates that the photon is expected to be somewhere between the first BS and the
place where two detectors will be placed if we choose to detect the
photon behind the first BS. 

\begin{figure}
\begin{center}
\includegraphics[scale=1]{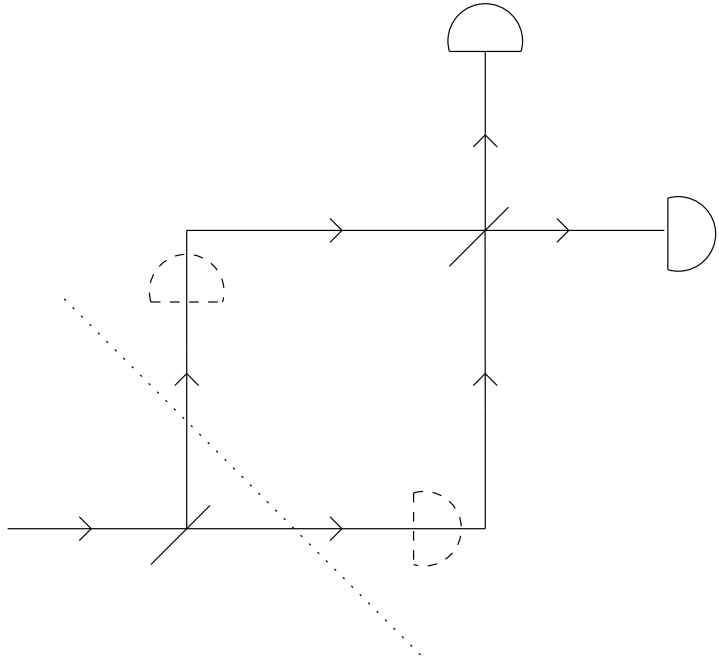}\\
\caption{Two identical 50/50 BS and two pairs of
  single-photon-detectors that can be placed either behind the first BS
  or the second. The dotted line
indicates that the photon is somewhere between the first BS and the
place where two detectors will be placed if we choose to detect the
photon behind the first BS.}
\end{center}
\end{figure}

What is the ``physical reality'' of
the photon at the stage indicated by the dotted line? In other
words, what and where is the photon at the stage indicated by the
dotted line?\\
Depending on whether we choose to detect the photon behind the first or the second
BS, we would be inclined to draw different conclusions on what is the nature of the photon
at this stage. But remember that we make
the choice between detecting the photon behind the first or second
BS only after the photon has been emitted by the source. Intuitively, one
would expect that a supposed objective ``state'' or ``objective configuration'' of the
photon would be completely determined by the source and the fact that we choose to detect
the photon behind the first or second BS does not in any way affect or
change its objective ``state'' at the stage
indicated by the dotted line. If this is the case, again it 
seems as if the photon can neither be of purely particle nature nor purely wave nature. 
But what is it then?\\ 
 
\section{A space-time approach to quantum mechanics}
\label{sec:astatqm}

In view of what have been said so far, let us consider the following possibility of
explaining the nature and behaviour of photons. Since photons upon detection behave
as though they were particles, maybe they are particles after all. If we
assume that they are particles, what else do we have to change then to
explain their un-particle-like behaviour? Let us go back to figure 4 again. 
One way in which we can uphold the idea that photons are particles, is if 
\textit{the path the photon takes depends on whether the detection is made
  behind the first or second BS}! In other words, \textit{the path
  the photon takes depends on the whole experimental setup}. 
This means for example that, which path the photon takes between the 
two BS:s depends not only on the source and the first BS, but also on 
whether the detectors are eventually placed behind the first or the
second BS.\\

Let us generalise this idea: Maybe the reason it is difficult to 
describe the objective configuration of the system between preparation 
and measurement, is that it is not determined by the preparation alone, but
by \text{both preparation and measurement} instead; i.e. knowing both what the prepared
state and the choice of measurement observable is. For example, 
assume that the state was prepared in $\ket{\!z+\!}$ and then measured in the 
$x$-direction. Then, in this situation we could assign the following
statistical distribution as a representative of the objective
statistical ``state'' of the system between preparation and measurement:
\begin{equation}
\{p_{(z+)(x+)},p_{(z+)(x-)}\} 
= 
\{|\braket{\!x\!+\!\!\!}{\!z\!+\!}|^2,|\braket{\!x\!\!-\!\!\!}{\!z\!+\!}|^2\}.  
\end{equation}
Here the two objective configurations of the system would be
influenced by the choice of measurement observable. So, by
giving up the idea that the preparation alone should completely
determine a ``state'' of the system, one has the possibility of
describing the system with \textit{one} statistical distribution over
some objective configurations.\\
 
Assume that this is really how things are. But how could it possibly be like
that!? How can a choice that is not yet made determine the objective
state of the photon now!? Does not this imply retrocausation,
i.e. future events being the cause of events in the past!?\\
 
Let us take the following view on the situation: In quantum 
mechanics the system is described by a set of statistical 
distributions, one for every possible measurement observable, 
which is called the quantum state of the system. The
theory is thus mathematically constructed so that the description of 
the system is formally independent of the settings (that determine
the measurement observable) of the measurement apparatus. This
reflects the belief that a pre-measurement description of the system
should be independent of the settings of the measurement apparatus and
this belief is put into the formal structure of quantum mechanics.
Viewed from an everyday or classical perspective this is a quite
natural assumption or construction. But taking into consideration the
strange behaviour of phenomena that quantum mechanics describes and
predicts, it is not necessarily a natural assumption to make anymore.
Also from a space-time perspective it is not necessarily a natural
assumption to make, as will be discussed more in what follows.\\

At the beginning of the twentieth century Einstein revolutionized
physics with his special theory of relativity \cite{einstein05}. 
The theory states that the speed of light is the highest signal
velocity; this fact is often
called ``Einstein causality''. The concept of space-time was
introduced and replaced the 
``Newtonian view''
on space and time as two separate entities. The concept of 
simultaneity no longer had any absolute
meaning. The theory suggests that one should see the world 
as a four-dimensional space-time that \textit{is} and not, as before,
as a three-dimensional absolute space that changes with time. By doing so
Einstein was later on able to understand gravity as curvature in this
space-time \cite{einstein16}. But although we can accept and intellectually understand
the concept of space-time, the human brain still perceives the world in a more
Newtonian way. We feel ``the passage of time'' and we only experience 
``one moment at the time''. This is probably a reason why
physicists still question whether the space-time should be
considered as something real or just a useful concept, despite the
success of the concept of space-time.\\

It is a common opinion (in view of Bell's theorem \cite{bell93}) that QM
is non-local. But the concept of non-locality is related to the
concept of simultaneity and things like action at a distance. In 
\cite{Einstein51}, p.61, Einstein writes:\\

\textit{There is no such thing as simultaneity of distant events; 
consequently there is also no such thing as immediate action at a
distance in the sense of Newtonian mechanics.}\\

There are several assumptions needed to prove Bell's
inequalities and from the fact that quantum mechanics does not 
satisfy Bell's inequalities one cannot draw the conclusion
that it is non-local (see e.g. \cite{morgan04,valdenebro02}). 
The assumptions needed to derive Bell's inequalities all seem
supported by common sense and dropping some of them seem to assume
strange ``non-local type correlations'' among instrument settings 
and hidden variables. But as Morgan writes in his paper \cite{morgan04}:\\

\textit{The violation of Bell inequalities can be modelled by entirely
local random fields, but leave an awkward question of how the nonlocal
correlations might have been established in the first place (that is,
how did the ``conspiracy'' arise?).}\\

So how could these type of correlations come about? What would be
their origin?\\ 
These type of correlations could perhaps find a more natural explanation when seen from a
space-time perspective. In the space-time view the whole universe is, 
so to say, created ``all at once''. The correlations would then be
there from the beginning. The space-time would be created as a
continuous whole, and correlations ``forward''
and ``backward'' in time would not be a conceptual problem when considering physical 
events. From a space-time perspective, a measurement ``forward'' in time would 
be similar to a preparation ``backward'' in time.\footnote{Similar ideas can be found in
  articles by O. Costa de Beauregard \cite{beauregard79,beauregard98}
  and references therein: ``A measurement thus is a reversed
  preparation - a retroparation. Such is the basis of the zigzagging
  causality model of EPR correlations.'' \cite{beauregard98}}
Here we have at least some motivation for the idea that an objective
pre-measurement ``state'' or configuration could depend not only on the preparation 
but also on the measurement apparatus settings.\\  

In cases such as classical mechanics, special relativity and 
normally for the hidden variable theories of QM, the theories are 
often assumed deterministic in the following sense. An exhaustive
description of all the variables on any time-slice of space and time, 
determines uniquely the variables for the whole space-time. When 
we say, in this paper, that the objective ``state'' of the system
should not only depend on the preparation but also on the measurement 
apparatus settings, we do not exclude the possibility that the
theory could be deterministic in the just mentioned sense. We are only
saying that maybe the origin of the ``non-local type correlations'' 
could find a more natural explanation when seen from a space-time 
perspective.\\

Let us again consider the experimental arrangement where an
apparatus prepares a ``spin up'' state in 
the z-direction and an apparatus will measure the 
state in some direction $\bar{n}$. The choice of $\bar{n}$ is
determined by an observer, e.g.\ a human or a machine. 
In spirit with the space-time approach, assume now 
that the experimental arrangement including the system
that chooses $\bar{n}$ is timelessly described and that the
measurement observable is some $\bar{n}^{\gamma}$.
Then the state between preparation and measurement could as before 
be seen as a normal statistical mixture of two ``objective configurations'':
\begin{equation}
\{p_{(z+)(\bar{n}^{\gamma}+)},p_{(z+)(\bar{n}^{\gamma}-)}\} 
= 
\{|\braket{\bar{n}^{\gamma}\!\!+\!\!\!}{\!z\!+\!}|^2,|\braket{\bar{n}^{\gamma}\!\!-\!\!\!}{\!z\!+\!}|^2\},  
\end{equation}
with probabilities as calculated from the rules of QM. Consider as another example
an arrangement that prepares the entangled Bell-state
\begin{equation}
\ket{\psi} \equiv \frac{1}{\sqrt{2}} (\ket{\!z+\!}\ket{\!z+\!}+\ket{\!z-\!}\ket{\!z-\!})
\end{equation}
and then measures the observables ($\bar{n}$, $\bar{n}'$) on the two
(spatially separated) subsystems. If we assume that
the (fixed) observables for this arrangement are 
($\bar{n}^{\delta}$,$\bar{n}^{\epsilon}$), the state 
between preparation and measurement could instead be seen as a normal 
statistical mixture of four ``objective configurations'':
\begin{equation}
\begin{split}
&\{p_{\psi(\bar{n}^{\delta}+)(\bar{n}^{\epsilon}+)}, p_{\psi(\bar{n}^{\delta}+)(\bar{n}^{\epsilon}-)},
  p_{\psi(\bar{n}^{\delta}-)(\bar{n}^{\epsilon}+)}, p_{\psi(\bar{n}^{\delta}-)(\bar{n}^{\epsilon}-)}\} =\\ 
&\{|\bra{\!\bar{n}^{\delta}\!\!\!+\!\!}\bra{\!\bar{n}^{\epsilon}\!\!\!+\!\!}\ket{\psi}|^2\!,   
  |\bra{\!\bar{n}^{\delta}\!\!\!+\!\!}\bra{\!\bar{n}^{\epsilon}\!\!\!-\!\!}\ket{\psi}|^2\!,
  |\bra{\!\bar{n}^{\delta}\!\!\!-\!\!}\bra{\!\bar{n}^{\epsilon}\!\!\!+\!\!}\ket{\psi}|^2\!,
  |\bra{\!\bar{n}^{\delta}\!\!\!-\!\!}\bra{\!\bar{n}^{\epsilon}\!\!\!-\!\!}\ket{\psi}|^2\}.  
\end{split}
\end{equation}

\section{``Space-time statistical mechanics''}
\label{sec:stsm}

\begin{figure}
\begin{center}
\includegraphics[scale=1]{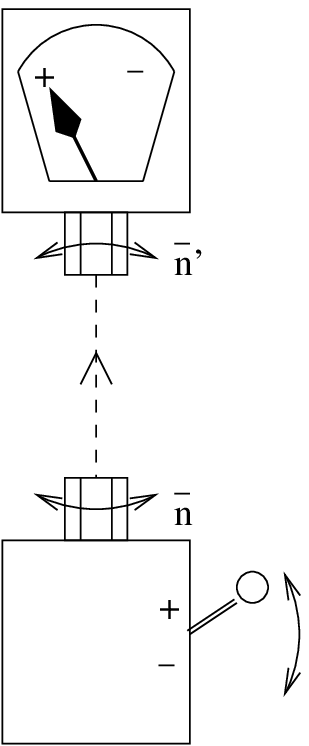}\\
\caption{A qubit is prepared in a state $\ket{\!\bar{n}+\!}$ by the lower
  apparatus. The upper apparatus then measures the state in the $\bar{n}'$
  direction and the outcome is $\bar{n}'\!+$. The macroscopic
  parameters of this arrangement are $\bar{n}+$ and $\bar{n}'$.}
\end{center}
\end{figure}

The considerations made so far give a hint that quantum mechanics maybe 
could be recreated or interpreted as a 
``space-time statistical mechanics'' (STSM). To explain what is meant
by that consider the following:
The settings on the preparation- and 
measurement apparatus are under the control of the observer. He can 
e.g.\ choose what state he wants to prepare or what observable to
measure. These settings we call \textit{macroscopic parameters}. If
these parameters are fixed, one gets a specific experimental
arrangement (see figure 5). It could
e.g.\ be the arrangement where a state is prepared as ``spin-up''
in the $z$-direction and measured in the $x$-direction. The macroscopic
parameters would then be '$z+$' and '$x$'. The ``state'' of
the system between preparation and measurement corresponding to this 
arrangement is the statistical mixture of two objective configurations 
of the system:
\begin{equation}\label{eq6}
\{p_{(z+)(x+)},p_{(z+)(x-)}\} 
= \{|\braket{\!x\!\!+\!\!\!}{\!z\!+\!}|^2,|\braket{\!x\!\!-\!\!\!}{\!z\!+\!}|^2\},
\end{equation}
again with probabilities calculated from the rules of QM.
Which one of the two configurations is then realised and how should
one describe the system? One could treat this arrangement
as in classical statistical mechanics. 
We assume that the macroscopic space-time configuration of the arrangement
is determined by the macroscopic parameters. Further we assume that the
microscopic space-time configuration of the arrangement is 
completely determined by some \textit{microscopic
parameters}. The idea is then that there are many different
microscopic space-time configurations that give rise to a particular
macroscopic space-time configuration. Further, the microscopic parameters,
corresponding to the same macroscopic parameters, also determine 
which of the two configurations in \eqref{eq6} that is realized and thus
the measurement outcome. Considering an ensemble of identical
experimental arrangements, all with the same macroscopic 
parameters and where by arrangement we mean a preparation device together with a 
measurement apparatus (and a possible observer), one would have a representation of the 
statistical mixture in \eqref{eq6}. The method presented here is 
analogous to the statistical mechanical method, 
which is the reason why we talk about ``space-time statistical
mechanics''. To be able to recreate quantum mechanics as a ``space-time 
statistical mechanics'', one would probably need to find rules for how
the microscopic space-time configurations are ``drawn'' and
``selected'', and then find a systematic way of treating
statistical ensembles over microscopic space-time configurations 
that gives rise to the probabilities that can be calculated from QM.\\

\section{The STSM idea applied in quantum state estimation}
\label{sec:QSE}

Let us apply the STSM idea in quantum state estimation. \cite{jaynes57,jaynes62,MMB05,shorejohnson80}
Consider $\textit{many}$ spin-1/2 systems each in some unknown state
$\ket{\theta_k,\phi_k}$, where $\theta_k \in [0,\pi]$ and $\phi_k \in
[0,2\pi[$ for $k=1,2,3,...$,  and assume that from many experiments we have 
observed that the mean value of the observable 
$\hat{\sigma_z}$ is $\bar{\sigma_z}$. To every $\ket{\theta_k,\phi_k}$
corresponds a vector $\bar{n}(\theta_k,\phi_k)$ on the Bloch sphere for
which 

\begin{equation}
(\bar{n}(\theta_k,\phi_k) \cdot (\hat{\sigma}_x,\hat{\sigma}_y,\hat{\sigma}_z))
\ket{\theta_k,\phi_k} = \ket{\theta_k,\phi_k}.
\end{equation}

According to the STSM idea, the set $S$ of
possible objective configurations of the system can be labeled by
$(\bar{n}(\theta,\phi)+,\bar{n}(0,0)\pm)$ or for short $(\theta,\phi,r)$, where
$\bar{n}(0,0)\cdot (\hat{\sigma}_x,\hat{\sigma}_y,\hat{\sigma}_z)
= \hat{\sigma}_z$ and $r = \pm 1$.
The systems are distributed over the configurations
$(\theta,\phi,r)$ with the probability density $p(\theta,\phi,r)$.\\
How can we choose the most unbiased $p(\theta,\phi,r)$ which is
compatible with the data $\bar{\sigma_z}$? -- By means of the maximum-relative-entropy
method. \cite{jaynes57,jaynes62,MMB05,shorejohnson80} But 
$p(\theta,\phi,r) = p(r|\theta,\phi) \, p(\theta,\phi)$ and since we want 
$p(r|\theta,\phi)$ to obey the rules of quantum mechanics,
i.e. $p(+1|\theta,\phi) = \cos^2{(\frac{\theta}{2})} \equiv q(+1|\theta,\phi)$ and
$p(-1|\theta,\phi) = \sin^2{(\frac{\theta}{2})} \equiv
q(-1|\theta,\phi)$, we are not going to use the Lagrange-multiplier method
with this non-linear constraint. Instead we apply the 
maximum-relative-entropy method on probability densities of the form 
$\tilde{p}(\theta,\phi,r) \equiv q(r|\theta,\phi) \, p(\theta,\phi)$,
not varying $q(r|\theta,\phi)$ but only $p(\theta,\phi)$.\\

First, since the mean value comes from many experiments, we can assume 
that the preparation must be such that

\begin{equation}\label{eqDC}\begin{split}
\bar{\sigma_z} = \langle\hat{\sigma}_z\rangle 
&\equiv \sum_r \int_{\theta=0}^{\pi}\int_{\phi=0}^{2\pi}
r \cdot \tilde{p}(\theta,\phi,r) \sin{\theta} \, d\phi \, d\theta\\ 
&= \sum_r \int_{\theta=0}^{\pi}\int_{\phi=0}^{2\pi}
r \cdot q(r|\theta,\phi) p(\theta,\phi) \sin{\theta} \, d\phi \, d\theta\\ 
&= \int_{\theta=0}^{\pi}\int_{\phi=0}^{2\pi}
[q(+1|\theta,\phi) - q(-1|\theta,\phi)]\, p(\theta,\phi) \sin{\theta} \,
d\phi \, d\theta\\ 
&= \int_{\theta=0}^{\pi}\int_{\phi=0}^{2\pi}
[\cos^2(\theta/2) - \sin^2(\theta/2)]\, p(\theta,\phi) \sin{\theta} \,
d\phi \, d\theta\\ 
&= \int_{\theta=0}^{\pi}\int_{\phi=0}^{2\pi}
\cos\theta\, p(\theta,\phi) \sin{\theta} \,
d\phi \, d\theta.
\end{split}\end{equation}

This constraint of course does not select a unique $\tilde{p}(\theta,\phi,r)$,
but we can choose the one among them which has maximum relative
entropy

\begin{equation}\label{eqSh}\begin{split}
H^{(3)}(\tilde{p}(\theta,\phi,r)) &\equiv - \sum_r \int_{\theta=0}^{\pi}\int_{\phi=0}^{2\pi}
\tilde{p}(\theta,\phi,r) \,log \frac{\tilde{p}(\theta,\phi,r)}{\tilde{m}(\theta,\phi,r)}
\sin{\theta} \, d\phi \, d\theta\\ 
&= \int_{\theta=0}^{\pi}\int_{\phi=0}^{2\pi}
p(\theta,\phi) \,log \frac{p(\theta,\phi)}{m(\theta,\phi)}
\sin{\theta} \, d\phi \, d\theta \equiv H^{(2)}(p(\theta,\phi)).
\end{split}\end{equation}

Here $\tilde{m}(\theta,\phi,r)$ is the prior probability density and 
$\tilde{m}(\theta,\phi,r) = q(r|\theta,\phi) \, m(\theta,\phi)$, since the prior probability
distribution should also obey the rules of quantum mechanics. In the last
step we have used that $p(+1|\theta,\phi)+p(-1|\theta,\phi)=1$.
We see from \eqref{eqSh} that, to find the $\tilde{p}(\theta,\phi,r)$ that 
maximizes $H^{(3)}(\tilde{p}(\theta,\phi,r))$, it is enough to find the $p(\theta,\phi)$
that maximizes $H^{(2)}(p(\theta,\phi))$ and the sought probability density
will then be given by $\tilde{p}(\theta,\phi,r) = q(r|\theta,\phi) \, p(\theta,\phi)$.
Let $p'(\theta,\phi)$ be the probability density that maximizes
$H^{(2)}(p(\theta,\phi))$ under the data constraint in \eqref{eqDC},
where the prior probability density $m(\theta,\phi)$ is taken to be
constant; $m(\theta,\phi) = 1/(4\pi)$. The solution to the non-linear
constraint problem is 

\begin{equation}
\tilde{p}'(\theta,\phi,r) = q(r|\theta,\phi) \,
\frac{m(\theta,\phi)}{Z} \, e^{-\lambda \cos{\theta}}
\end{equation}

where

\begin{equation}
Z \equiv \int_{\theta=0}^{\pi}\int_{\phi=0}^{2\pi}
m(\theta,\phi) \, e^{-\lambda \cos{\theta}} \sin{\theta} \, d\phi \, d\theta.
\end{equation}

(The Lagrange multiplier $\lambda$ is determined from the data
constraint.) One can easily show that to this probability density 
corresponds the density matrix 

\begin{equation}
\hat{\rho} = \sum_{r}\int_{\theta=0}^{\pi}\int_{\phi=0}^{2\pi} \tilde{p}'(\theta,\phi,r)\,
\ket{\theta,\phi}\bra{\theta,\phi}\,\sin\theta\, d\phi\, d\theta 
= \frac{1}{2}\,(\hat{I}+\bar{\sigma}_z \,\hat{\sigma_z}).
\end{equation}

Note that this is the same density matrix one would find with the 
``von Neumann entropy method'' \cite{jaynes57,jaynes62}, in
which one instead seeks the density matrix that maximizes the von 
Neumann entropy, under the same data constraint 
$\bar{\sigma_z} =\, \langle\hat{\sigma}_z\rangle$. 

\begin{acknowledgements}
The author would like to thank Piero G. Luca Mana, Peter Morgan,
Anders Karlsson and Gunnar Bj{\"o}rk for advice, encouragement, and 
useful discussions.
\end{acknowledgements}

\bibliography{andersbib}

\end{document}